\documentclass[conference]{IEEEtran}
\IEEEoverridecommandlockouts
\usepackage{bbding}
\usepackage{float}  
\usepackage{subfig}
\usepackage{graphicx}
\usepackage{subcaption} 
\usepackage{booktabs}
\usepackage{siunitx}
\usepackage{graphicx} 

\usepackage{cite}
\usepackage[hyphens]{url}
\usepackage[breaklinks, colorlinks, linkcolor=purple, anchorcolor=purple, citecolor=purple, urlcolor=purple]{hyperref}

\usepackage{amsmath,amssymb,amsfonts}
\usepackage{algorithmic}
\usepackage{graphicx}
\usepackage{textcomp}
\usepackage{xcolor}

\newtheorem{remark}{Remark}
\def\BibTeX{{\rm B\kern-.05em{\sc i\kern-.025em b}\kern-.08em
    T\kern-.1667em\lower.7ex\hbox{E}\kern-.125emX}}

\begin{document}
\title{Two-Stage Stochastic Optimization for Low-Carbon Dispatch in a Combined Energy System}

\author{\IEEEauthorblockN{Manling Hu}
\IEEEauthorblockA{Department of Earth \& Environmental Science\\
University of Pennsylvania\\
Philadelphia, USA\\
huml@sas.upenn.edu}
\and
\IEEEauthorblockN{ Manqi Xu}
\IEEEauthorblockA{ Tsinghua-Berkeley Shenzhen Institute\\
Tsinghua University\\
 Shenzhen, China\\
 xmq22@mails.tsinghua.edu.cn}
\and
\IEEEauthorblockN{Dunnan Liu*}
\IEEEauthorblockA{School of Economics \\and Management\\
North China Electric Power University\\
Beijing, China\\
liudunnan@163.com}
\thanks{Corresponding author: Dunnan Liu, e-mail: liudunnan@163.com. This work was partly supported by Natural Science Foundation of China (No. 72171082).}
}

\maketitle

\begin{abstract}
While wind and solar power contribute to sustainability, their intermittent nature poses challenges when integrated into the grid. To mitigate these issues, renewable energy can be combined with coal-fired power and hydropower sources to stabilize the energy system, with battery storage serving as a backup source to smooth the total output. This study develops a low-carbon dispatch model for a combined energy system using a two-stage stochastic optimization approach. The model incorporates a carbon trading mechanism to regulate emissions and addresses the uncertainty in wind and solar outputs by clustering output curves into typical scenarios to derive a joint distribution. In the initial stage of scheduling, decisions are made regarding the unit commitment for coal-fired power plants. The second stage optimizes the expected operation cost of other energy generation sources. The feasibility of the model is demonstrated by comparing the results of stochastic and deterministic scenarios through simulation. Analysis of different carbon prices further explores the impact of the carbon trading mechanism on the system’s operation cost.
\end{abstract}


\section{Introduction}
As traditional fossil fuels become depleted and renewable energy experiences unprecedented growth, the electric power system is shifting toward a cleaner, more efficient, and economically sustainable energy framework. In this era of energy transition, high emissions from existing coal-fired power plants (CFPPs) present a significant challenge to achieving the 1.5°C climate goal, while the intermittency and variability of renewable energy sources create obstacles to grid reliability. Relying exclusively on either traditional or renewable energy sources is not a viable long-term solution. Instead, an integrated combined energy system harnesses the complementary strengths of diverse generation sources, offering a more resilient approach.

In a combined energy system, the coordinated output from each generation source is key to achieving efficient planning and operation. Traditionally, deterministic optimization methods have been used, where all inputs are assumed to be known and fixed. However, these methods often fail to capture the inherent uncertainty in renewable outputs. To address this issue, advancements in generation unit commitment (UC) and economic dispatch (ED) models under uncertainty have been made. These improvements include robust optimization, distributionally robust optimization, and stochastic optimization \cite{ref1}. Robust optimization aims to minimize the highest possible cost considering all potential scenarios with uncertain inputs. For example, an adaptive robust optimization model is proposed in the coordinated operation of wind, solar, and battery storage to address the UC problem\cite{ref2}. Considering the combination of traditional coal power units and wind power, a distributionally robust planning model for UC problems is developed in \cite{ref3}. Although robust optimization reduces the need for scenario-based modeling, it can lead to higher operational costs due to its conservative nature.

Stochastic optimization considers the minimization of the expected cost across all scenarios. In unit commitment and economic dispatch (UCED) problems, it accounts for both the fixed cost of scheduling generating units and the variable cost of dispatching them based on uncertainties. A security-constrained unit commitment stochastic model is proposed to combine wind and traditional power systems, utilizing neural network-based prediction intervals to address wind power forecast uncertainties \cite{ref4}. A stochastic programming model is proposed for the planning of a multi-source energy system comprising hydropower, thermal power, wind power, and pumped storage systems \cite{ref5}. A multistage stochastic programming approach is employed to manage microgrids that are operated with variable renewable energy sources and battery storage, taking into account short- and long-term uncertainties\cite{ref6}. Scenario-based stochastic optimization is widely regarded as a common approach for capturing uncertainty, relying on scenario representation algorithms to model various uncertainty factors. A scenario mapping technology is introduced in a stochastic unit commitment model to compress a large number of renewable energy scenarios \cite{ref7}.

Traditional energy system optimization typically focuses on minimizing operational costs, often neglecting the environmental impact. As the energy transition accelerates, there is an increasing focus on low-carbon power dispatch, which considers the external costs of greenhouse gas (GHG) emissions and the interaction of electricity and carbon markets. A distributed robust optimization model is developed for the wind and thermal power combined system in the carbon market, using carbon emission reductions and cost savings as the key evaluation criteria \cite{ref8}. A risk-constrained ED model is introduced for electricity-gas systems, incorporating carbon trading prices and integrating carbon capture systems and demand response technologies \cite{ref9}. In the integrated energy systems (IES), the stochastic scheduling model is often combined with the segmented index-based carbon market model \cite{ref10} and life cycle assessments \cite{ref11} to evaluate and limit greenhouse gas emissions. 

The contributions of this paper can be summarized as follows: (1) a two-stage stochastic programming model is proposed to address the uncertainties in renewable energy sources for UCED problem, and (2) a carbon trading mechanism is integrated to enable low-carbon dispatch.
The rest of the paper is organized as follows. Section II introduces the framework of the proposed low-carbon dispatch model and the carbon trading mechanism. Section III establishes the two-stage stochastic optimization model for the combined energy system. In Section IV, a case study and analysis of the impact of various carbon prices are conducted. Conclusions are drawn in Section V.

\section{Framework}
\subsection{Overview}
\begin{figure}[h]
    \centering
    \includegraphics[width=0.8\linewidth]{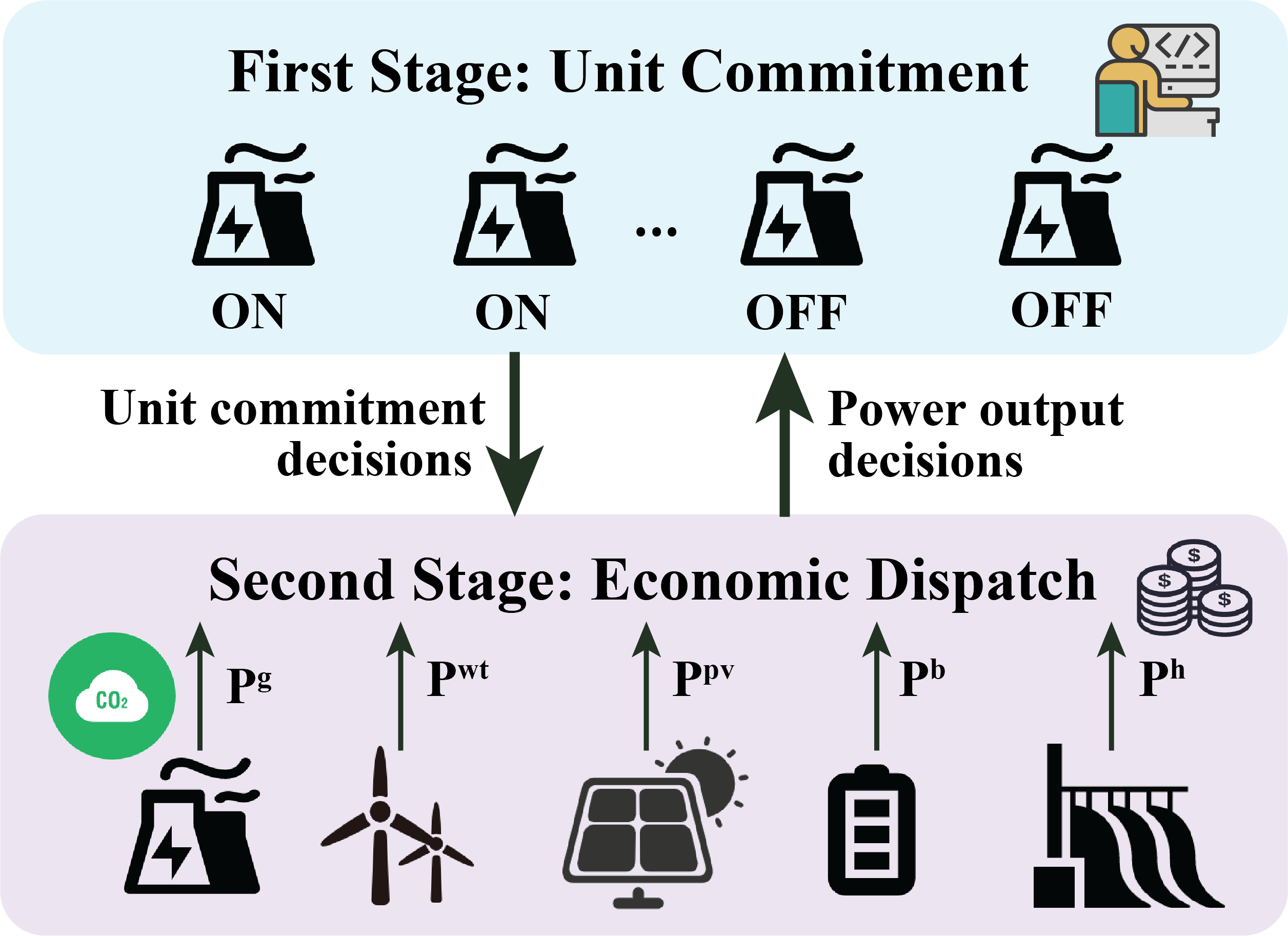}
    \caption{Two-Stage UCED problem structure}
    \label{fig:overview}
\end{figure}

In this paper, we formulate the low-carbon dispatch problem as a two-stage stochastic problem, as shown in Fig. \ref{fig:overview}. 
The first stage focuses on the UC decisions for CFPP units in the day-ahead market. 
The second stage involves ED across various renewable energy scenarios in the intraday market. The goal is to minimize the expected system operating cost, consisting of the fuel cost and carbon emission cost.
\subsection{Carbon trading mechanism}
To effectively reduce CO$_2$ emissions from the energy system, it is crucial to integrate renewable energy with market-based mechanisms. Under the carbon trading mechanism, CFPPs are incentivized to phase out older, inefficient units and adopt advanced technologies such as carbon capture and storage.  Additionally, coupling with the electricity market, it provides incentives for renewable energy generation and further encourages renewable energy generation by increasing the marginal cost of CFPPs, thus fostering a clean energy transition \cite{ref12,ref13}.

In the carbon market, the traded commodity is mainly the $\text{CO}_2$  emission allowance, which sets a limit on the total $\text{CO}_2$ emissions for each emitter. The initial common allowance is distributed for free in the global CO$_2$ emission trading market. The CO$_2$ emission cost structure of CFPPs is as follows:
 \begin{align}
C^{\text{CO}_2} &= \lambda^{\text{CO}_2} \left( E_{\text{CO}_2} - E_0 \right) \notag \\
&= \lambda^{\text{CO}_2} \left\{ \sum_{t\in \mathcal{T}} \sum_{i\in \mathcal{N}} E_i(P^g_{i,t}) - E_0 \right\} \label{3},
\end{align}
where $\mathcal{N}=\{1,....,N\}$ is the set of indices for CFPPs, and $\mathcal{T}=\{1,....,T\}$ is the set of dispatch period. $E_{\text{CO}_2}$ is the total CO$_2$  emissions of the units, $\lambda^{\text{CO}_2}$ is the carbon trading price, and $E_i(P^g_{i,t})$ is the CO$_2$  emissions of unit $i$ at time $t$ for power generation $P^g_{i,t}$. The initial allocation of CO$_2$  emissions $E_0$ is derived based on the relationship between carbon emissions, as shown in equation (2):

\begin{equation}
E_0 =  \sum_{t=1}^{T} \sum_{i=1}^{N} \eta h P^g_{i,t},
\end{equation}
where $\eta$ is the correction factor for the adjustment of free allocation. $h$ is the allocation coefficient for carbon emissions per unit of electricity generation. 

$E_i(P^g_{i,t})$ is calculated as follows:

\begin{equation}
E_i(P^g_{i,t}) = l_i + k_i P^g_{i,t} + j_i (P^g_{i,t})^2.
\end{equation}
The function $E_i(P^g_{i,t})$ is determined by the reduction coefficients $l_i$, $k_i$, and $j_i$ for each unit and the amount of electricity generated.

\section{Two-stage Stochastic Optimization Model}

\subsection{First-stage problem: unit commitment}\label{AA}
The first stage problem models the unit commitment problem where the objective is to minimize the start-up and shut-down costs of the CFPP units, along with the expected system operation cost under each typical scenario. 
The objective function is shown as follows:

\begin{equation}
    \min_{u_{i,t}} \mathcal{F} = \sum_{t\in \mathcal{T}} \sum_{i\in \mathcal{N}} C^{uc}_{i,t} + \mathbb{E}_{\omega}[h(u, \omega)], \quad 
\end{equation}
where $\mathcal{F}$ is the total cost composed of two parts:
\begin{itemize}
\item $C^{uc}_{i,t}$ represents the unit commitment cost of unit $i$ at time $t$, which is calculated as follows:
\begin{equation}
  C^{uc}_{i,t}= u_{i,t} (1 - u_{i,t-1}) SU_i + u_{i,t-1} (1 - u_{i,t}) SD_i,
\end{equation}
where $u_{i,t}$ is a binary variable indicating whether the unit $i$ is on or off at time $t$ (0 for off, 1 for on), $SU_i$ and $SD_i$ are the start-up cost and the shut-down cost for unit $i$, respectively.
\item  $\mathbb{E}_{\mathbb{P}}[h(u, \omega)]$ is the expected system economic dispatch cost for all scenarios $\omega$:
\begin{align}
    \mathbb{E}_{\mathbb{P}}[h(u, \omega)]=\sum_{\omega \in \Omega} \pi_{\omega} h(u, \omega),
\end{align}
where $\Omega$ is the set of typical scenarios obtained through wind and solar power output clustering. $\pi_{\omega}$ is the occurrence probability of scenario $\omega$, and $h(u, \omega) $ is explained in the (\ref{2-obj}).

\end{itemize}

The decision variable of the first stage is the unit commitment decision $\{u_{i,t}, \forall i \in \mathcal{N}, \forall t \in \mathcal{T}\}$. The unit commitment constraints are described as follows:
\subsubsection{Unit start-up and shut-down time constraints}

The frequency of units starting up and shutting down affects the CFPP fuel cost. Therefore, the following constraints are imposed on the continuous start-up and shut-down times:
\vspace{-5pt}
\begin{align}
    & \forall i \in \mathcal{N}, \forall t \in \mathcal{T}:\notag\\
    & \sum_{k=t}^{t+TS-1} \left( 1 - u_{i,k} \right) \geq TS \left( u_{i,t-1} - u_{i,t} \right),\\
    & \sum_{k=t}^{t+TO-1} u_{i,k} \geq TO \left( u_{i,t} - u_{i,t-1} \right), 
\end{align}
where $TS$ and $TO$ represent the minimum continuous start-up and shut-down times for the unit.
\subsubsection{Unit start-up and shut-down cost constraints}

\begin{align}
& C_{i,t}^{U} \geq H_i \left( u_{i,t} - u_{i,t-1} \right),C_{i,t}^{U} \geq 0, \forall t \in \mathcal{T},  \\
& C_{i,t}^{D} \geq J_i \left( u_{i,t-1} - u_{i,t} \right), C_{i,t}^{D} \geq 0, \forall t \in \mathcal{T}, 
\end{align}
where $H_i$ and $J_i$ represent the one-time start-up cost and shut-down cost of unit $i$, and $C_{i,t}^{U}$ and $C_{i,t}^{D}$ represent the start-up and shut-down costs, respectively.

\subsection{Second stage problem: economic dispatch}
In the second stage, we aim to optimize the expected operating cost, considering the impact of carbon trading. Wind, solar, and hydro generation do not incur fuel costs, so in the second stage, the ED cost only includes the fuel costs of CFPP units. The objective function for each scenario $\omega$ is given by:
\begin{equation}
h(u, \omega)=\min_{x_\omega} \quad C^{\text{fuel}}_\omega+C^{\text{CO}_2}_\omega,\label{2-obj}
\end{equation}
where $C^{\text{fuel}}_\omega$ is the fuel cost of the CFPP generation, and $C^{\text{CO}_2}_\omega$ is the cost of carbon emission, which is calculated based on the carbon trading mechanism (\ref{3}).
$C_{\text{fuel}}$ for each scenario $\omega$ is given in (\ref{13}):
\begin{align}
    &C^{\text{fuel}}_\omega = \sum_{t\in\mathcal{T}} \sum_{i\in \mathcal{N}} \lambda^{\text{coal}} u_{i,t} f(P^g_{\omega,i,t}),\label{13}\\
    & f_i(P^g_{\omega,i,t}) = a_i + b_i P^g_{\omega,i,t} + c_i (P^g_{\omega,i,t})^2,\label{14}
\end{align}
where $\lambda^{\text{coal}}$ is the purchasing price of coal converted into standard coal, $f(P_{i,t,\omega})$ is the fuel consumption function of unit $i$ in scenario $\omega$ as defined in (\ref{14}), which is determined by the power generation $P_{i,t,\omega}$. $a_i$, $b_i$, and $c_i$ are fuel cost parameters determined by the capacity and characteristics of unit $i$.

The decision variable of the second stage is $x_{\omega}=\{ P^g_{\omega,i,t},P^{\text{wt}}_{\omega,t},P^{\text{pv}}_{\omega,t}, P^h_{\omega,t},P^{b}_{\omega,t},E_{\omega,t}, \forall t\in \mathcal{T}\}$. The economic dispatch should be subject to the following constraints:
\subsubsection{System power balance constraint}

Neglecting electricity losses, the total output of wind, solar, CFPP, hydropower, and battery storage sources must match the system load curve at any time. The power balance constraint of the multi-source energy system is given by:
\begin{align}
    \sum_{i\in \mathcal{N}} P^g_{\omega,i,t} + P^{\text{wt}}_{\omega,t} + P^{\text{pv}}_{\omega,t} + P^h_{\omega,t}+ P^{b}_{\omega,t} = P^l_t,\forall t\in \mathcal{T},
\end{align}
where $\sum_{i\in \mathcal{N}} P^g_{\omega,i,t}$ is the total output of all CFPP units in scenario $\omega$ at time $t$, $P^{\text{wt}}_{\omega,t}$ is the real-time output of the wind farm, $P^{\text{wt}}_{\omega,t}$ is the real-time output of the solar power plant, $P^h_{\omega,t}$ is the real-time output of the hydropower, $P^{b}_{\omega,t}$ is the net discharging power of the battery storage system, and $P^l_t$ is the electricity load at time $t$.
\subsubsection{CFPP generation constraint}
\begin{equation}
u_{i,t} P^{g,-}_{i} \leq P^g_{\omega,i,t} \leq u_{i,t} P^{g,+}_{i},  \forall i \in \mathcal{N}, \forall t \in \mathcal{T},
\end{equation}
where $P^{g,+}_{i}$ and $ P^{g,-}_{i}$ are the maximum and minimum output of the CFPP unit $i$, respectively.

\subsubsection{CFPP ramp rate constraint}
\begin{equation}
- R_d \leq P^g_{\omega,i,t} - P^g_{\omega,i,t-1}\leq R_u, \forall i \in \mathcal{N}, \forall t \in \mathcal{T},
\end{equation}
where $R_d$ is the ramp-down rate limit, and $R_u$ is the ramp-up rate limit.

When the minimum output $P^{g,-}_{i}$ of the unit during start-up exceeds the ramp-up rate $R^u$, the constraint will prevent all previously shut-down units from starting. Therefore, it can be rewritten as:
\begin{align}
   & \forall i \in \mathcal{N}, \forall t \in \mathcal{T}:\notag\\
   & P^g_{\omega,i,t} - P^g_{\omega,i,t-1} \leq u_{i,t-1}(R^u - S_{i}^u) + S_{i}^u,\\
   & P^g_{\omega,i,t} - P^g_{\omega,i,t-1} \geq u_{i,t-1}(R^d - S_{i}^d) + S_{i}^d, 
\end{align}
where $S_{i}^u$ and $S_{i}^u$ are the maximum ramp-up and ramp-down rates of unit $i$. For simplification, we define:

\begin{equation}
S_{i}^u = S_{i}^u= \frac{1}{2}(P_{i}^{g,-} + P_{i}^{g,+})
\end{equation}

\subsubsection{Wind farm output constraint}
\begin{equation}
0 \leq P^{\text{wt}}_{\omega,t} \leq \bar{P}^{\text{wt}}_{\omega,t}, \forall t \in \mathcal{T},
\end{equation}
where $\bar{P}^{\text{wt}}_{\omega,t}$ is the maximum output of the wind farm at time $t$ in scenario $\omega$.

\subsubsection{Solar power plant output constraint}

\begin{equation}
0 \leq P^{\text{pv}}_{\omega,t} \leq \bar{P}^{\text{pv}}_{\omega,t}, \forall t \in \mathcal{T},
\end{equation}
where $\bar{P}^{\text{pv}}_{\omega,t}$ is the maximum output of the solar power plant at time $t$ in scenario $\omega$.

\subsubsection{Hydropower plant output constraint}

\begin{equation}
0 \leq P^h_{\omega,t} \leq \bar{P}^h_{\omega,t}, \forall t \in \mathcal{T},
\end{equation}
where $\bar{P}^h_{\omega,t}$ is the maximum output of the hydropower plant at time $t$ at scenario $\omega$.

\subsubsection{Battery storage constraints}

Since the internal circuit of the battery does not need to be considered, the energy balance and charging/releasing power constraints for the battery storage model are established as follows, from equation (\ref{5-1}) to equation (\ref{5-5}):
\begin{align}
 & P^b_{\omega,t} = P^{re}_{\omega,t} - P^{ch}_{\omega,t},\forall t \in \mathcal{T},\label{5-1}\\
 &0 \leq P^{ch}_{\omega,t} \leq  P^{ch,+}_{\omega,t}, \forall t \in \mathcal{T}, \label{5-2}\\
 & 0 \leq P^{re}_{\omega,t} \leq  P^{re,+}_{\omega,t}, \forall t \in \mathcal{T}, \label{5-3}\\
 & P^{re}_{\omega,t}\perp  P^{ch}_{\omega,t}, \forall t \in \mathcal{T},\label{nosimu}\\
 & E_{\omega,t+1} = E_{\omega,t} +  P^{ch}_{\omega,t}\eta_c \Delta t - \frac{P^{re}_{\omega,t} \Delta t}{\eta_d},\forall \leq T-1,\label{energy_update}\\
 & SoC^{min} B \leq E_{\omega,t} \leq  SoC^{\text{max}} B,  \forall t \in \mathcal{T},\label{5-5}
\end{align}
where $P^{ch}_{\omega,t}$ and $P^{re}_{\omega,t}$ are the electricity charging and releasing rates in scenario $\omega$ at time $t$. 
Constraints (\ref{5-2})-(\ref{5-3}) limit the charging and releasing power of the battery storage system, where $P^{ch,+}_{\omega,t}$ and $P^{re,+}_{\omega,t}$ are the maximum charging and releasing limits. 
Constraint (\ref{nosimu}) ensures that simultaneous charging and releasing are forbidden.
Constraint (\ref{energy_update}) defines the battery energy update. $E_{\omega,t}$ represent the stored energy level at time $t$ for scenario $\omega$, $\eta_c$ and $\eta_d$ are the charging and releasing efficiency rates of the battery storage system, respectively.
Constraint (\ref{5-5}) limits the storage energy level within the safe operation bound, where $SoC^{min}$ and $SoC^{max}$ represent the minimum and maximum states of charge (SoC) for the system, $B$ is the rated system capacity.
\begin{remark}
    The proposed two-stage stochastic UCED problem is formulated as a Mixed Integer Quadratic Programming problem, where non-linearity arises from the fuel cost and carbon cost components. We reformulate it as a  Mixed Integer Linear Programming using piecewise linearization methods \cite{lin2013review}, and it is omitted here for brevity.
\end{remark}

\section{Case Study}
The proposed low-carbon stochastic dispatch model is tested on a combined energy system, consisting of CFPP units, a wind farm, a hydropower plant, a solar power plant, and a battery storage system. The model is tested on a personal computer using MATLAB R2022b, YALMIP, and GUROBI 9.1.2.

\subsection{Data construction}\label{ITH}
\begin{figure}[h!]
    \centering
    \includegraphics[width=0.65\linewidth]{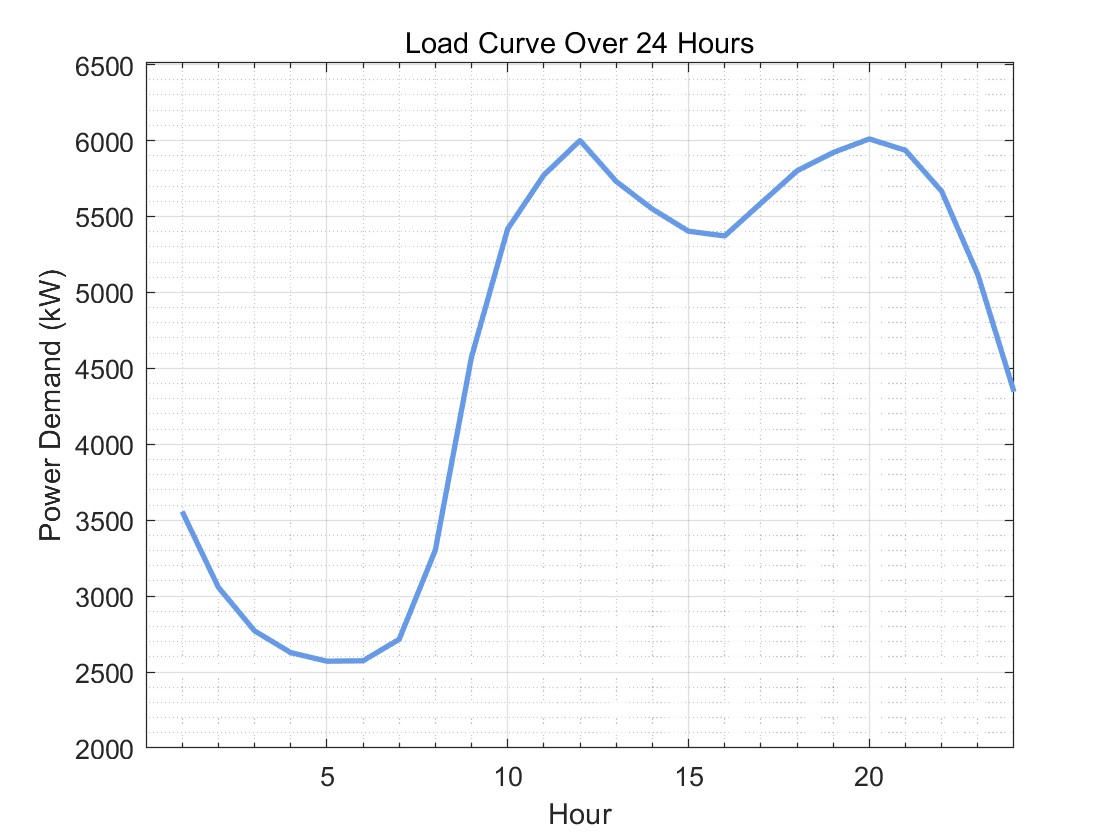}
    \caption{Load curve over 24 hours}
    \label{figure3}
    \vspace{-10pt}
\end{figure}

The capacity of the CFPP is 3070MW, the hydropower plant is 100MW, the wind farm is 1250MW, the solar power plant is 300MW, and the battery energy system is 600MW. The battery has a power limit of 80 MW, with state of charge limits ranging from 0.3 to 0.9. It operates with a charging efficiency of 0.95 and a releasing efficiency of 1/0.95. The tested load shown in Fig. \ref{figure3}, wind, and solar power outputs are based on 48 days of data from a region in Hubei Province, China. For the carbon market, we set the carbon trading price to be 100RMB/ton,  $\eta$ to be 1 and $h$ to be 0.9419. The scheduling time interval is one hour.
\subsection{Scenario analysis}\label{ITH}
Scenario analysis is critical in solving the two-stage stochastic UCED problem, given the inherent variability of renewable energy outputs. This paper employs k-means clustering to reduce the dimensionality of time-series data while preserving the original dataset’s statistical properties. The algorithm converges toward representative scenarios by iteratively recalculating cluster centers based on Euclidean distance. The optimal number of clusters is determined using the Elbow method, minimizing the sum of squared errors to balance computational efficiency and data fidelity.

\begin{figure}[h!]
    \centering
    \subfloat[Wind power output for 3 scenarios]{\includegraphics[width=0.495\columnwidth]{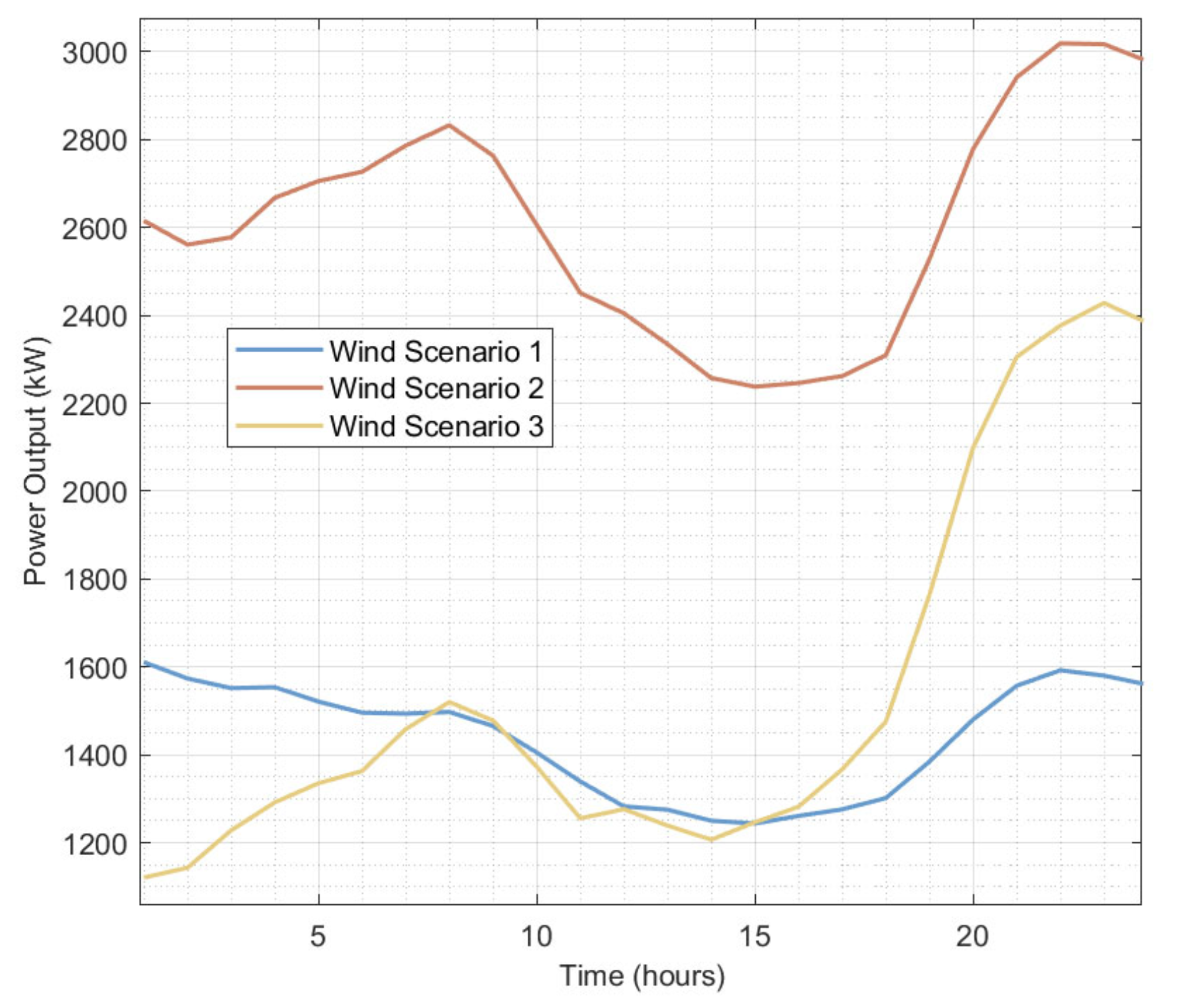}}%
    \label{figure}
    \hfill
    \subfloat[Solar power output for 3 scenarios]{\includegraphics[width=0.48\columnwidth]{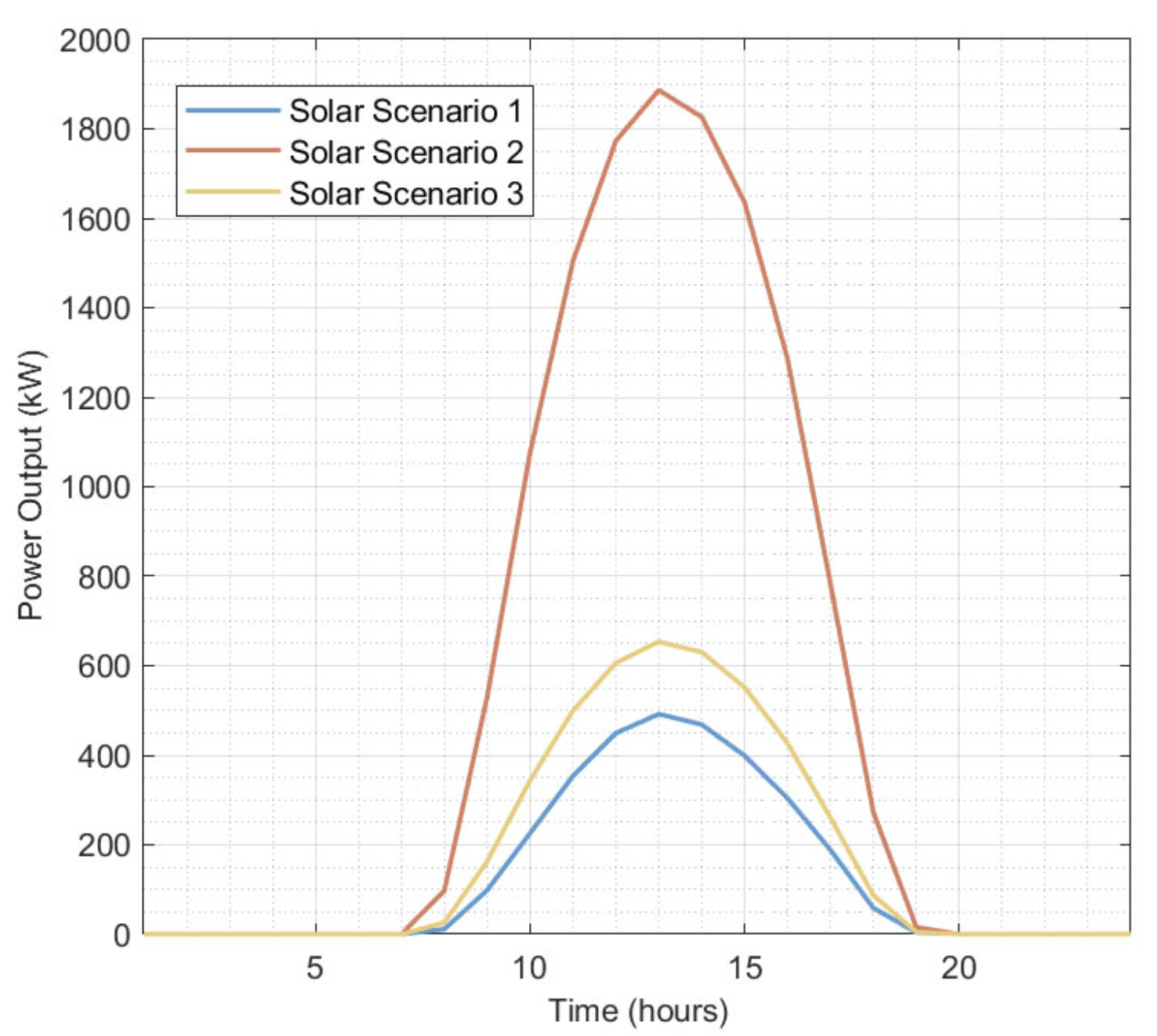}}%
    \label{figure}
    \caption{Renewable energy output scenarios}
    \label{figure2}
\end{figure}

For the case study, 48 days of wind and solar power output data are reduced to three representative scenarios respectively using k-means clustering, as shown in Fig. \ref{figure2}. Nine distinct scenarios with associated probabilities—44.40\%, 21.48\%, 2.86\%, 9.42\%, 4.56\%, 0.61\%, 10.76\%, 5.21\%, and 0.70\%—are obtained after applying a Cartesian product. These scenarios serve as input for the stochastic optimization process, effectively capturing renewable generation variability.

\subsection{Simulation results}\label{FAT}
\begin{figure}[h]
    \centering
    \includegraphics[width=0.9\linewidth]{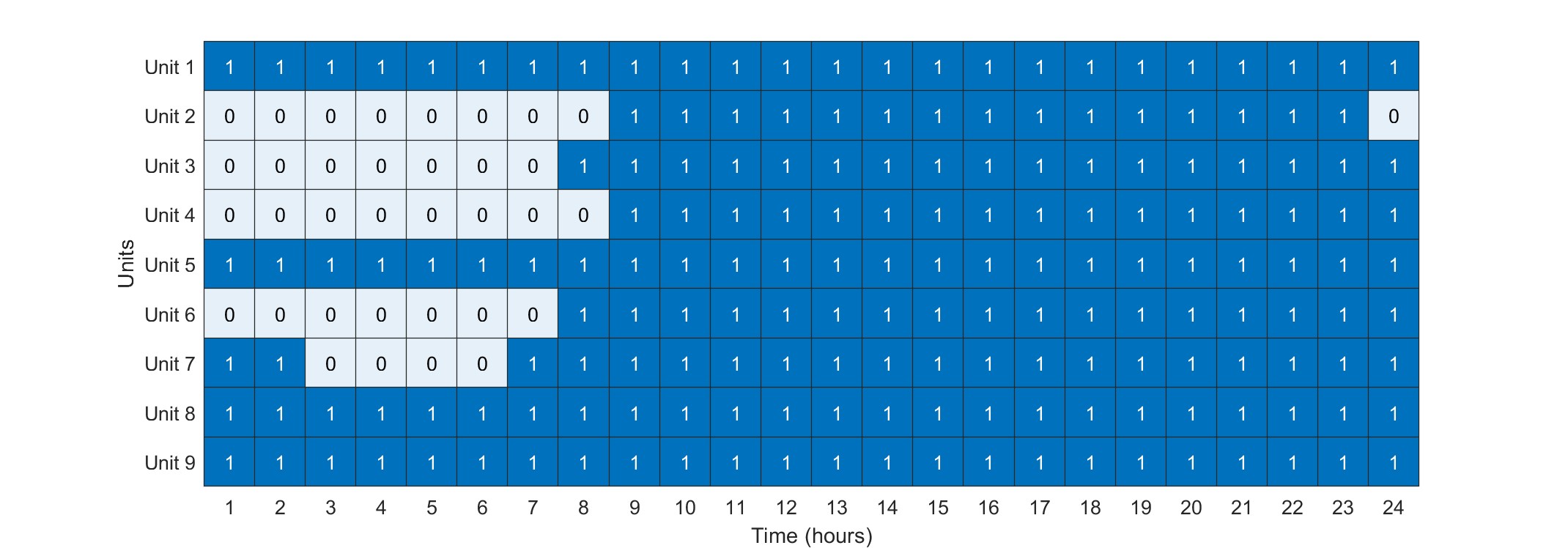}
    \caption{Unit commitment decisions (0 = Off, 1 = On)}
    \label{figure4}
\end{figure}

\begin{figure}[h]
    \centering
    \includegraphics[width=\linewidth]{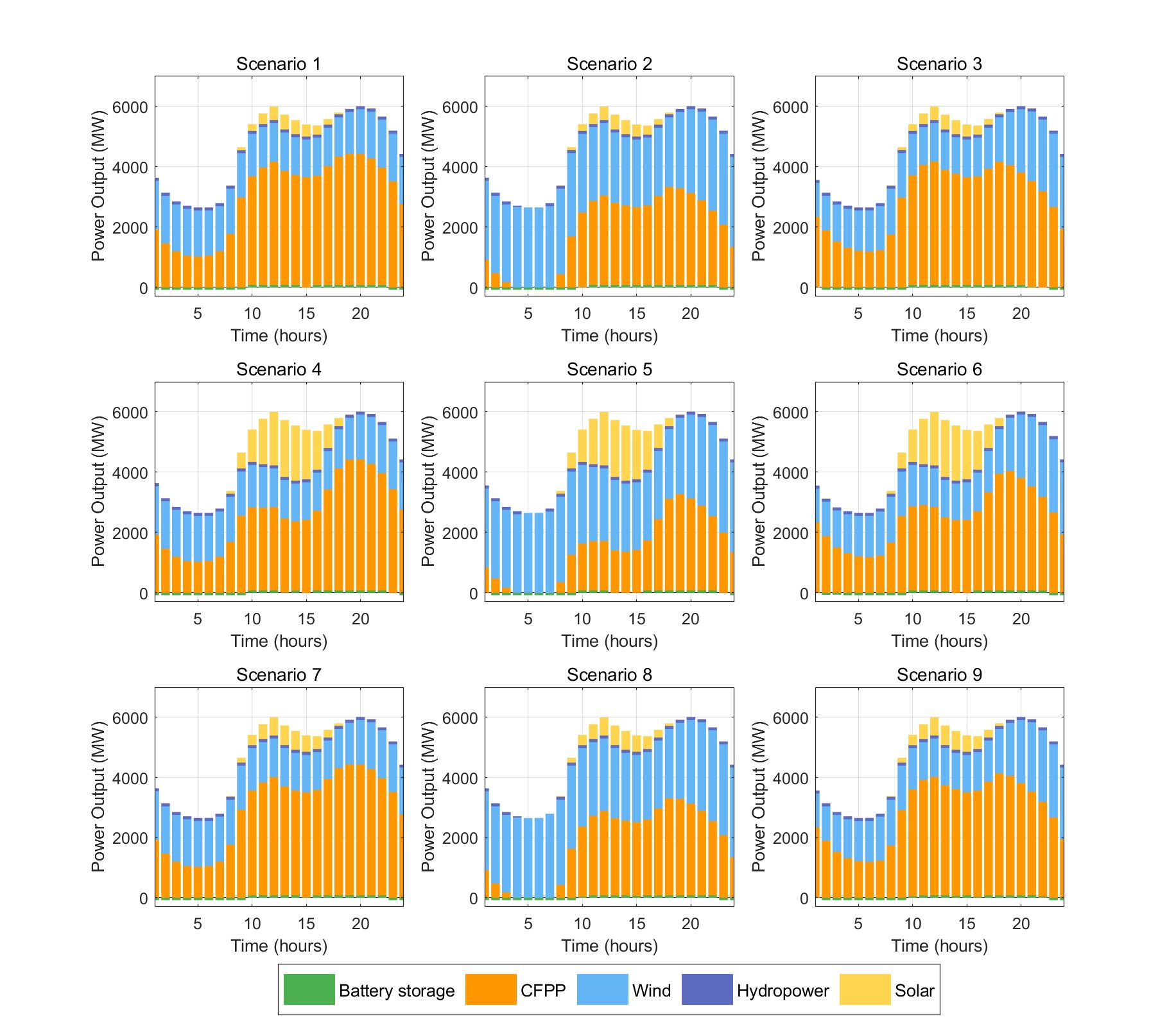}
    \caption{Economic dispatch results of nine scenarios }
    \label{figure5}
\end{figure}

Taking scenario probabilities into account, Fig. \ref{figure4} illustrates the Unit commitment status for CFPP units in the first stage of stochastic optimization. Due to the high start-up costs, units 2, 3, 4, and 6 remain offline during the early morning hours when demand is relatively low. The results of the low-carbon ED for the nine scenarios are presented in Fig. \ref{figure5}.

To compare the total cost, start-up/shut-down costs, and start-up/shut-down cycles between deterministic and stochastic optimization, a deterministic optimization for a typical day is performed without considering the uncertainties in wind and solar output. The comparison results are presented in Table \ref{DOSO}, demonstrating that stochastic optimization outperforms deterministic optimization. It reduces the total cost and lowers start-up/shut-down times, indicating better operational stability and reduced wear on CFPP units.

\begin{table}[h]
    \centering
    \caption{Comparison of stochastic and deterministic optimization}
     \label{DOSO}
    \resizebox{\linewidth}{!}{ 
    \begin{tabular}{lS[table-format=1.4]S[table-format=5.0]S[table-format=2.0]}
        \toprule
        \textbf{Optimization Method} & \textbf{Total Cost (e+10)} & \textbf{Cycling Cost} & \textbf{Start-Stop Cycle} \\
        \midrule
        Stochastic Optimization      & 3.9111 & 82000 & 7 \\
        Deterministic Optimization   & 4.5006 & 102000 & 10 \\
        \bottomrule
    \end{tabular}
    } 
\end{table}

Various carbon prices are applied in the model to evaluate their impact on the proposed stochastic optimization model, as shown in Table \ref{table:carbon_price_impact}. The carbon prices are set at 0, 42.85, 100, 130, and 300 RMB/ton, respectively.

\begin{table}[h]
    \centering
    \caption{Carbon price impact on stochastic optimization}
    \label{table:carbon_price_impact}
    \resizebox{0.7\linewidth}{!}{ 
    \begin{tabular}{c c S[table-format=3.3e1]}
        \toprule
        \textbf{Case} & \textbf{Carbon Price (RMB/ton)} & \textbf{Total Cost (RMB)} \\
        \midrule
        1 & 0      & 1.304e9  \\
        2 & 42.85  & 1.767e10 \\
        3 & 100    & 3.911e10 \\
        4 & 130    & 5.036e10 \\
        5 & 300    & 1.141e11 \\
        \bottomrule
    \end{tabular}
    } 
\end{table}

\section{Conclusion}
This paper developed a combined energy system optimization approach based on renewable energy scenario analysis and carbon trading mechanism. A two-stage stochastic optimization model accounting for CFPP unit commitment planning and economic dispatch of other clean energy sources is constructed and solved. 

The case study demonstrates that compared with deterministic optimization, stochastic optimization can improve both cost efficiency and operational flexibility by accounting for uncertainties in renewable energy generation. It also analyzes the effects of increasing carbon prices, showing that the carbon trading mechanism has a direct impact on the operation cost of the combined energy system.






\bibliographystyle{IEEEtran}
\bibliography{IEEEabrv,BibTeX}

\begin{thebibliography}{10}
\providecommand{\url}[1]{#1}
\csname url@rmstyle\endcsname
\providecommand{\newblock}{\relax}
\providecommand{\bibinfo}[2]{#2}
\providecommand\BIBentrySTDinterwordspacing{\spaceskip=0pt\relax}
\providecommand\BIBentryALTinterwordstretchfactor{4}
\providecommand\BIBentryALTinterwordspacing{\spaceskip=\fontdimen2\font plus
\BIBentryALTinterwordstretchfactor\fontdimen3\font minus \fontdimen4\font\relax}
\providecommand\BIBforeignlanguage[2]{{%
\expandafter\ifx\csname l@#1\endcsname\relax
\typeout{** WARNING: IEEEtran.bst: No hyphenation pattern has been}%
\typeout{** loaded for the language `#1'. Using the pattern for}%
\typeout{** the default language instead.}%
\else
\language=\csname l@#1\endcsname
\fi
#2}}

\bibitem{ref1}
L.~A. Roald, D.~Pozo, A.~Papavasiliou, D.~K. Molzahn, J.~Kazempour, and A.~Conejo, ``Power systems optimization under uncertainty: {A} review of methods and applications,'' \emph{Electric Power Systems Research}, vol. 214, p. 108725, Jan. 2023.

\bibitem{ref2}
A.~Lorca and X.~A. Sun, ``Multistage robust unit commitment with dynamic uncertainty sets and energy storage,'' \emph{IEEE Transactions on Power Systems}, vol.~32, no.~3, pp. 1678--1688, May 2017.

\bibitem{ref3}
P.~Xiong, P.~Jirutitijaroen, and C.~Singh, ``A distributionally robust optimization model for unit commitment considering uncertain wind power generation,'' \emph{IEEE Transactions on Power Systems}, vol.~32, no.~1, pp. 39--49, Jan. 2017.

\bibitem{ref4}
H.~Quan, D.~Srinivasan, and A.~Khosravi, ``Incorporating wind power forecast uncertainties into stochastic unit commitment using neural network-based prediction intervals,'' \emph{IEEE Transactions on Neural Networks and Learning Systems}, vol.~26, no.~9, pp. 2123--2135, Sept. 2015.

\bibitem{ref5}
M.~Daneshvar, B.~Mohammadi-Ivatloo, K.~Zare, and S.~Asadi, ``Two-stage stochastic programming model for optimal scheduling of the wind-thermal-hydropower-pumped storage system considering the flexibility assessment,'' \emph{Energy}, vol. 193, p. 116657, Feb. 2020.

\bibitem{ref6}
P.~Aaslid, M.~Korpås, M.~M. Belsnes, and O.~B. Fosso, ``Stochastic optimization of microgrid operation with renewable generation and energy storages,'' \emph{IEEE Transactions on Sustainable Energy}, vol.~13, no.~3, pp. 1481--1491, July 2022.

\bibitem{ref7}
E.~Du, N.~Zhang, C.~Kang, and Q.~Xia, ``Scenario {Map} {Based} {Stochastic} {Unit} {Commitment},'' \emph{IEEE Transactions on Power Systems}, vol.~33, no.~5, pp. 4694--4705, Sept. 2018.

\bibitem{ref8}
J.~Jin, Q.~Wen, S.~Cheng, Y.~Qiu, X.~Zhang, and X.~Guo, ``Optimization of carbon emission reduction paths in the low-carbon power dispatching process,'' \emph{Renewable Energy}, vol. 188, pp. 425--436, Apr. 2022.

\bibitem{ref9}
Y.~Xiang, G.~Wu, X.~Shen, Y.~Ma, J.~Gou, W.~Xu, and J.~Liu, ``Low-carbon economic dispatch of electricity-gas systems,'' \emph{Energy}, vol. 226, p. 120267, July 2021.

\bibitem{ref10}
L.~Wang, Z.~Shi, W.~Dai, L.~Zhu, X.~Wang, H.~Cong, T.~Shi, and Q.~Liu, ``Two-stage stochastic planning for integrated energy systems accounting for carbon trading price uncertainty,'' \emph{International Journal of Electrical Power \& Energy Systems}, vol. 143, p. 108452, Dec. 2022.

\bibitem{ref11}
R.~Wang, X.~Wen, X.~Wang, Y.~Fu, and Y.~Zhang, ``Low carbon optimal operation of integrated energy system based on carbon capture technology, {LCA} carbon emissions and ladder-type carbon trading,'' \emph{Applied Energy}, vol. 311, p. 118664, Apr. 2022.

\bibitem{ref12}
H.-r. Wang, T.-t. Feng, and C.~Zhong, ``Effectiveness of {CO2} cost pass-through to electricity prices under “electricity-carbon” market coupling in china,'' \emph{Energy}, vol. 266, p. 126387, Mar. 2023.

\bibitem{ref13}
J.~Wu, Y.~Fan, G.~Timilsina, Y.~Xia, and R.~Guo, ``\BIBforeignlanguage{en}{Understanding the economic impact of interacting carbon pricing and renewable energy policy in {China}},'' \emph{\BIBforeignlanguage{en}{Regional Environmental Change}}, vol.~20, no.~3, p.~74, June 2020.

\bibitem{lin2013review}
M.-H. Lin, J.~G. Carlsson, D.~Ge, J.~Shi, and J.-F. Tsai, ``A review of piecewise linearization methods,'' \emph{Mathematical problems in Engineering}, vol. 2013, no.~1, p. 101376, 2013.

\end{thebibliography}

\end{document}